\documentclass[journal=nalefd,manuscript=article]{achemso}

\usepackage{chemformula} 
\usepackage[T1]{fontenc} 
\usepackage{siunitx}
\usepackage{color}

\author{V.\ Cleric\`{o}}
\affiliation{Group of Nanotechnology, USAL-NANOLAB,  Universidad de Salamanca, E-37008 Salamanca, Spain}
\author{J.\ A.\ Delgado-Notario}
\affiliation{Group of Nanotechnology, USAL-NANOLAB,  Universidad de Salamanca, E-37008 Salamanca, Spain}
\author{M.\ Saiz-Bret\'{\i}n}
\affiliation{Departamento de F\'{\i}sica de Materiales, Universidad Complutense, E-28040 Madrid, Spain}
\author{A.\ V.\ Malyshev}
\affiliation{Departamento de F\'{\i}sica de Materiales, Universidad Complutense, E-28040 Madrid, Spain}
\author{Y.\ M.\ Meziani}
\affiliation{Group of Nanotechnology, USAL-NANOLAB,  Universidad de Salamanca, E-37008 Salamanca, Spain}
\author{P.\ Hidalgo}
\affiliation{Departamento de F\'{\i}sica de Materiales, Universidad Complutense, E-28040 Madrid, Spain}
\author{B. M\'{e}ndez}
\affiliation{Departamento de F\'{\i}sica de Materiales, Universidad Complutense, E-28040 Madrid, Spain}
\author{M.\ Amado}
\affiliation{Group of Nanotechnology, USAL-NANOLAB,  Universidad de Salamanca, E-37008 Salamanca, Spain}
\author{F.\ Dom\'{\i}nguez-Adame}
\affiliation{Departamento de F\'{\i}sica de Materiales, Universidad Complutense, E-28040 Madrid, Spain}
\author{E.\ Diez}
\affiliation{Group of Nanotechnology, USAL-NANOLAB,  Universidad de Salamanca, E-37008 Salamanca, Spain}
\email{enrisa@usal.es}
\phone{+34 670 581 543}
\fax{+34 923 29 48 53}

\title{Quantum nanoconstrictions fabricated by cryo-etching in encapsulated graphene}


\begin{document}








\begin{abstract}

More than a decade after the discovery of graphene, ballistic transport in nanostructures based on this intriguing material still represents a challenging field of research in two-dimensional electronics. The presence of rough edges in nanostructures based on this material prevents the appearance of truly ballistic electron transport as theo\-re\-tically predicted and, therefore, not well-developed plateaus of conductance have been revealed to date. In this work we report on a novel implementation of the cryo-etching method, which enabled us to fabricate graphene nanoconstrictions encapsulated between hexagonal boron nitride thin films with unprecedented control of the structure edges.  High quality smooth nanometer-rough edges are characterized by atomic force microscopy and a clear correlation between low roughness and the existence of well-developed quantized conductance steps with the concomitant occurrence of ballistic transport is found at low temperature. In par\-ti\-cu\-lar, we come upon exact 2$e^{2}/h$ quantization steps of conductance at zero magnetic field due to size quantization, as it has been theoretically predicted for truly ballistic electron transport through graphene nanoconstrictions. 

\end{abstract}


The use of two-dimensional (2D) materials for electronic-like devices blossomed after the discovery of graphene and represents an exciting whole field of research on its own. In such structures, the extreme existing aspect ratio, where the width $W$ and length $L$ of the devices exceed orders of magnitude the thickness, and the importance of manufacturing the devices with low-roughness borders, play a crucial role in the nature of the electronic transport. Graphene stands out as one of the most intriguing 2D materials in which the possibility of achieving a fully-tunable electronic confinement was pursued from its discovery\cite{Han}. In the ba\-llis\-tic regime, graphene is able to sustain quantized conductance with electron transport levels equally spaced in energy\cite{VanWees}, but not prior to overcome some issues such as the intrinsically moderate mobility in the single-layer-graphene samples and the importance of the borders in the mechanically exfoliated sheet of carbon atoms that can completely change the bulk conductance within the sample.\cite{CaridadNL} To overcome these problems that are detrimental for the study of pure ballistic transport, heterostructures of graphene encapsulated within thin layers of diverse 2D materials such as hexaboron-nitride (hBN)\cite{Kretinin} have been manufactured, obtaining mobilities exceeding orders of magnitude those of the standalone single-layer-graphene structures. However, the problem of the roughness at the borders remains. 

Single-layer and encapsulated-like graphene samples, where the width of the conductive layer has been narrowed down to few hundreds or eventually some tens of nanometers, have served as the template for the study of ballistic transport of carriers. To date, there have been two independent ways of achieving quantized conductance in graphene, either using electrostatic potential gates placed on the top (underneath) the flake to confine the density of charge into a well defined area or by tailoring physically the two-dimensional flake into narrow structures, \emph{i.e.} into nanoconstrictions~(NCs), through a mechanical or chemical etching definition. While the first approach has shown a promising outcome with well defined conductance plateaus separated $2e^{2}/h$ and $4e^{2}/h$\cite{Kim,Overweg}, the latter has turned into an arduous technological challenge with no clear evidence of quantized plateaus. In the first mechanically-defined graphene NCs deposited directly onto silicon dioxide wafers\cite{Han,Lu}, the presence of a plethora of different transport effects such as Coulomb blockade, high electron backscattering and electron localization by disorder and charging effects, together with an intrinsically low mobility of carriers, resulted in conductance steps at least an order of magnitude lower than the conductance quantum $2e^{2}/h$\cite{Lin,Lian}. Some years ago, Tombros \emph{et al.}\cite{Niko} observed $2e^{2}/h$ steps of conductance in a large mobility constriction on suspended graphene defined by current annealing. The nanostructure was build by passing a high current through a previously suspended graphene flake, stretching and deforming it into a NC-like structure. The main limitation of this method is purely technological since the accurate control of the dimensions of the graphene NC is unattainable. 

As previously stated, the use of lattice-matched hBN layers was a great leap forward into high-quality nanostructures, allowing for the fabrication of much higher mobility devices with intercalated graphene\cite{Bischoff}. Electron transport properties of samples of encapsulated graphene show signatures of quantum conductance in form of \emph{kinks}\cite{Terres,Terres2}. Yet, the effect of roughness of the structure due to the intrinsic dielectric behavior of the hBN film and the lack of control in the edge-definition of the heterostructures (of paramount importance to reduce edge scattering) have been detrimental for the appearance of well-defined plateaus of conductance. The problem of edge-definition and high roughness values was partially overcome thanks to the definition of GNCs with exfoliated graphene on hydrophobic silicon oxi\-de substrate with the use of hexamethyldisilazane (HMDS)\cite{Caridad,Vito}. This method increases the graphene mobility for a fair period of time ($\sim \SI{96}{\hour}$) after its treatment. The results presented in Refs.~[\hspace*{-4px}\citenum{Caridad,Vito}] shown quantized conductance at zero magnetic field that appears, once more, in form of \emph{kinks}. This procedure represents a promising compromise between relatively high mobility and very good edge-definition (roughness of the order of $\SI{1}{\nano\meter}$), but improvements on the etching definition in higher mobility and more stable graphene GNCs are most desirable.

In this work we present the fabrication method and characterization of NCs on encapsulated graphene where a cryo-etching technique commonly applied for bulk three-dimensional materials has been used in quasi-2D structures achieving an unprecedented control of the edge-definition. This technique was successfully implemented in silicon-based devices, where it was possible to control and define low-roughness sidewalls, observe the absence of sidewall scalloping\cite{Defforgea}, and generate samples with a high aspect ratio and cleaning\cite{Cadarso} in nanometric-size structures\cite{LiuZ}. The extension of the cryo-etching technique to graphene allowed us to built NCs with very low sidewall roughness, as proved by atomic force microscopy (AFM) measurements. AFM micrographs were employed to assess the actual profile of the sample edges, that were later used as input in numerical simulations. We found an excellent agreement between the measured and calculated  electrical conductivity and a good correspondence to the theoretical prediction as well.

We focus the attention on a hBN/graphene/hBN constriction of a few hundreds of $\SI{}{\nano\meter}$ width $W$. The fabrication process started from heterostructures of hBN/graphene/hBN, obtained by dry transfer technique in a way very similar to the one reported by Pizzocchero \emph{et al.}\cite{Pizzocchero}. An area free of impurities and bubbles in the hBN was selected by inspecting the different heterostructures with an optical Raman confocal microscope ($\lambda=\SI{532}{\nano\meter}$, $600$\,g/mm grating and spot size $\SI{1}{\micro\meter}$, see supplemental material S1). A first electron beam lithography (EBL) and a dry etching process in induced coupled plasma with SF$_6$ atmosphere were subsequently performed to define a bar-like structure in the hBN/graphene/hBN heterostructure. A rapid thermal annealing in Ar atmosphere at $\SI{350}{\celsius}$ for $15$ minutes removed eventual small non-visible bubbles. With a second self-aligned EBL step and the electron beam evaporation of Cr/Au ($\SI{5/45}{\nano\meter}$) we defined and deposited side contacts for electric measurements. 

The constrictions were defined by means of the innovative cryo-etching process in gra\-phene-based structures in the third aligned EBL step, which was performed at $\SI{-110}{\celsius}$ in a controlled Ar/SF$_6$ ($10/40$\,sccm) atmosphere (see supplemental information S2). The use of a  polymethylmethacrylate (PMMA) mask for this etching process represents a less invasive approach for encapsulated graphene-structures compared to the use of a Cr mask in submicron-size constrictions.\cite{Terres} To the best of our knowledge, this is the first time that a cryo-etching method is successfully used for the definition of graphene nanostructures. 

\begin{figure}[ht]
\centering{\includegraphics[width=0.95\columnwidth,clip=]{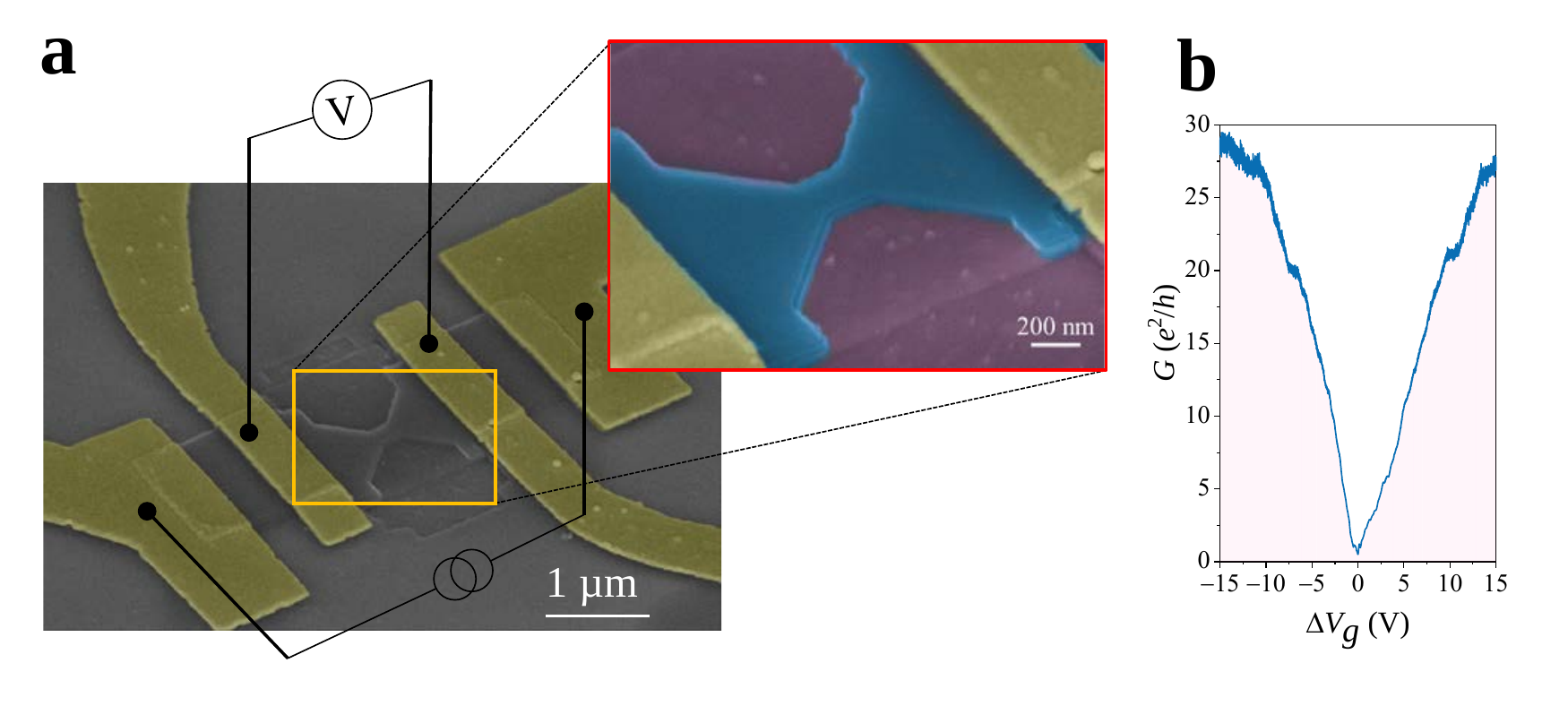}}
\caption{(a) Tilted SEM micrograph of an encapsulated graphene NC with lateral width $W\simeq \SI{206}{\nano\meter}$ and length  $L\simeq \SI{200}{\nano\meter}$ including an schematic view of the electronic setup. The inset shows an enlarged view where the hBN/graphene/hBN heterostructure is colored in sky blue, the SiO$_{2}$ substrate (partially etched) in violet and the contacts in yellow gold. (b)~Conductance as a function of the normalized voltage $\Delta V_{g}=V_{g}-V^{*}_{g}$ for the same NC measured at $T=\SI{3.1}{\kelvin}$.}
\label{fig1}
\end{figure}

Figure~\ref{fig1}a shows a tilted SEM micrograph of a typical NC in encapsulated graphene defined by cryo-etching. The lateral width of this NC is $W\simeq \SI{206}{\nano\meter}$ and the length is $L\simeq \SI{200}{\nano\meter}$. The inset shows the SEM micrograph at higher magnification, in which the sandwiched structure (hBN/graphene/hBN) is colored in sky blue. A standard $4$--probes configuration to perform the electronic characterization at low-temperatures is sketched in the figure, where a pseudo-dc ($\SI{13}{\hertz}$) current of $\SI{5}{\nano\ampere}$ was injected/collected from the outer contacts while the voltage drop was measured by an in-phase lock-in preamplifier closer to the constriction (the contacts are gold colored for better clarity). The result of a typical measurement at $T=\SI{3.1}{\kelvin}$ is depicted in Figure~\ref{fig1}b. The conductance in units of $e^{2}/h$ is displayed as a function of the normalized back-gate voltage $\Delta V_{g}=V_{g}-V_{g}^{*}$, $V_{g}^{*}$ being the value of the voltage at the Dirac point. The conductance is directly obtained by the measured contactless resistance as $G=1/R$.  In our sample the hole-side region ($\Delta V_{g}<0$) is always better resolved than the electron one ($\Delta V_{g}>0$), where the ballistic behaviour starts at higher back-gate voltage due to the residual charge density. For this reason, in the main text we restrict our discussion to the hole-side (see supplementary information S4 for the electron-side regime). The typical mobility in our devices is $\mu\simeq \SI{150000}{\centi\square\meter/\volt\second}$ at room temperature, as shown in the supplementary information S3, obtained by means of the field effect formula\cite{Kim2}.

\begin{figure}[ht]
\centering{\includegraphics[width=0.9\columnwidth,clip=]{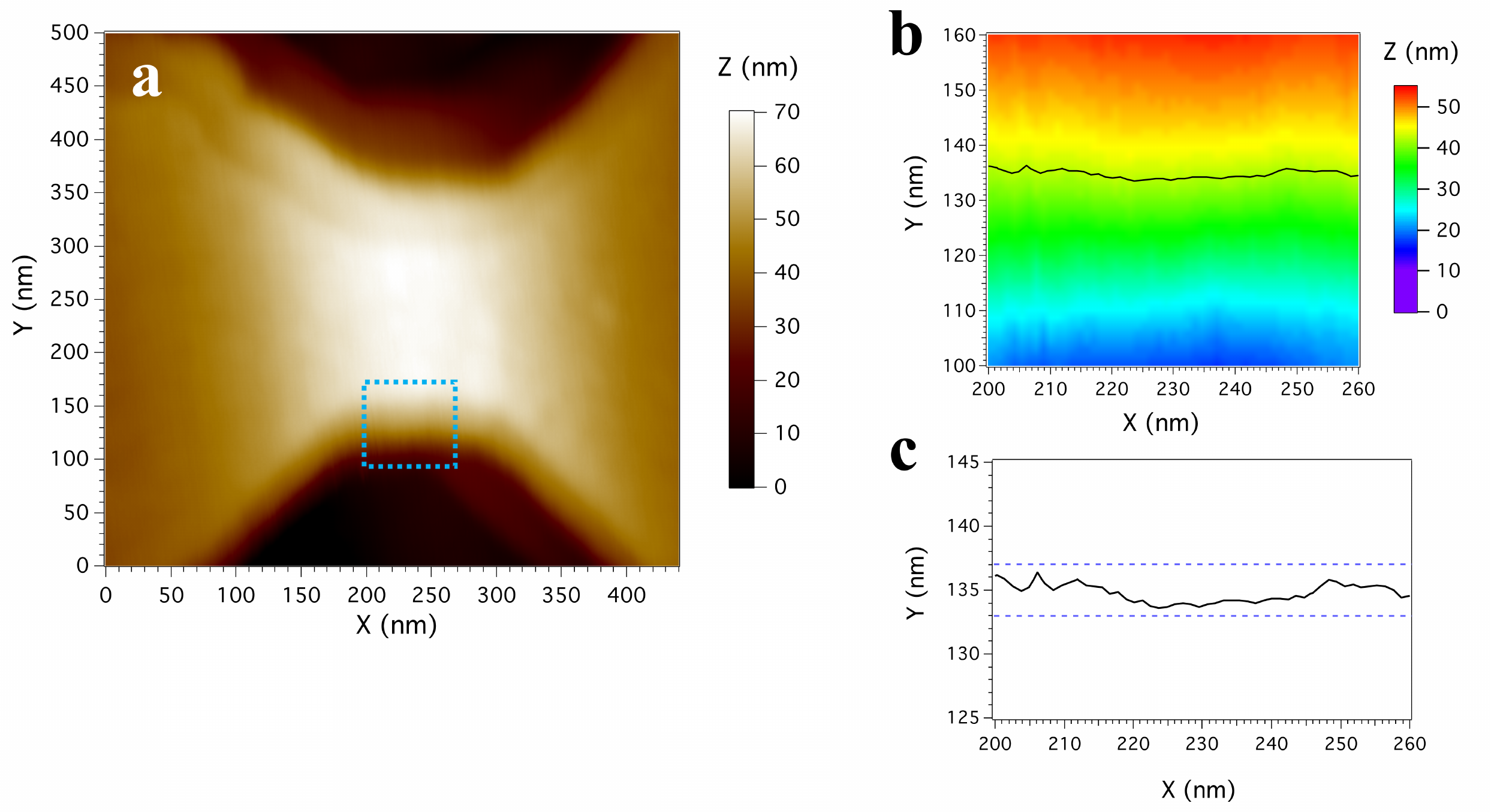}}
\caption{(a)~AFM image of the graphene NC with $W=\SI{206}{\nano\meter}$. (b)~Contour plot taken from the square highlighted in panel~(a), at $\SI{15}{\nano\meter}$ from the top of the nanostructure. (c)~Enlarged view of the contour plot, where the dashed lines indicate the values used to estimate the edge roughness.}
\label{fig2}
\end{figure}

The quality of the graphene NCs is apparent in the SEM micrograph shown in Figure~\ref{fig1}a. However, in order to get a quantitative analysis of the edge roughness, AFM measurements were performed in our samples. We have used a Nanotech AFM instrument operating in contact mode. Figure~\ref{fig2}a displays the AFM image of the whole NC. The edge and the corresponding contour profile within the square marked in panel~(a) are shown in Figure~\ref{fig2}b. 
The AFM images reveal the excellent definition of our graphene NC and, in particular, the smooth sidewall. The contour plot (black line in Figure~\ref{fig2}b) is taken $\SI{15}{\nano\meter}$ under the top hBN flake (with a thickness of about $\SI{15}{\nano\meter}$) in order to assess the actual size and roughness of the NC. The average edge roughness can be estimated by zooming in this profile and then a roughness around $\SI{2}{\nano\meter}$ is obtained (see Figure~\ref{fig2}c). This value is not far from the edge roughness estimated in lower mobility graphene NCs on HMDS\cite{Caridad}, namely without hBN, where the value was of the order of $\SI{1}{\nano\meter}$. Our measurement represents the first quantitative estimation of the roughness observed in encapsulated graphene nanoconstrictions as well.

Thanks to the accurate profile obtained with AFM measurements, we are able to study also the real edge roughness in our simulations based on a tight binding model. The electronic properties of graphene close to Dirac point can be accurately described by a tight-binding Hamiltonian for $\pi$ electrons,
%
%
$\mathcal{H}=-\sum_{\langle i,j\rangle}t_{ij}|i\rangle\langle j|$, where $|i \rangle$ is the atomic orbital of the $i$-th carbon atom~\cite{Wallace47}. The corresponding orbital energy level is set as the origin of energy without losing generality. Here $t_{ij}$ stands for the hopping energy parameter between orbitals of the $i$-th and $j$-th carbon atoms and has been set to be constant between nearest-neighbor atoms ($t_{ij}=2.8\,$eV, see Ref.~[\hspace*{-4px}\citenum{CastroNeto09}]) and zero otherwise. Such approximation provides reliable results near the $K$ and $K^{\prime}$ points of the Brillouin zone, as already reported in Ref.~[\hspace*{-4px}\citenum{Reich02}].

Neglecting electron-phonon~\cite{Morozov08,Milosevic10} and electron-electron\cite{Li10} interactions in our calculations, we consider electrons in the fully coherent regime, hence with ballistic motion through the whole device. Combining the \emph{quantum transmitting boundary method}, based on a finite element approximation~\cite{Lent90}, and an \emph{effective transfer matrix method} adapted for graphene (see Ref.~[\hspace*{-4px}\citenum{Munarriz14}] for technical details), the wave function in the whole sample and the transmission coefficient $\tau_n (E)$ for each mode at an energy $E$ is calculated. These modes, also known as channels or subbands, arise from the transverse quantization due to the lateral confinement at the leads~\cite{Datta95}. According to the Buttiker-Landauer formalism~\cite{Datta95}, at very low temperature the electrical conductance is proportional to the transmission coefficient at the Fermi energy~$E_F$
\begin{equation}
G=\frac{2e^2}{h}\sum_{n}\tau_{n}(E_F)\ ,
\end{equation}
where the index $n$ runs over the transverse modes. The factor $2$ accounts the spin degeneracy whereas the valleys are considered explicitly in our tight-binding calculation of the transmission coefficient $\tau_n$. As the NC becomes wider, the number of transverse modes for a given energy increases and so does the conductance. But interestingly, we found that the main features are essentially the same provided that the conductance is plotted against $Wk_F$, where $k_{F}$ is the Fermi wavenumber and $W$ is the NC width. This has an important practical consequence, namely we can extrapolate results to much larger systems as the ones fabricated experimentally and otherwise computationally very expensive (or even impossible) to simulate. 

In the low-temperature regime, when the elastic mean free path $l_{m}$ is larger than the GNC length $L=\SI{206}{\nano\meter}$,  transport is ballistic. We have estimated from the carrier mo\-bi\-li\-ty of our GNC that $l_{m}>\SI{1}{\micro\meter}$ (see supplementary information S3)  in the temperature range considered in this work. Hence conditions for ballistic transport are largely satisfied and therefore the conductance should be also described by the commonly used Sharwin formula\cite{Sharwin,Jong,Terres,Kumar} (replacing the usual factor $2$ by a factor $4$ to account for both spin and valley degeneracies in graphene) showing a linear dependence of $Wk_{F}$
\begin{equation}
G=\frac{4e^{2}}{\pi h} \, Wk_{F} t\ ,
\label{fit}
\end{equation}
where  $t$ is the transmission parameter, being $t=1$ for the ideal ballistic transport.

\begin{figure}[H]
\centering{\includegraphics[width=0.95\columnwidth,clip=]{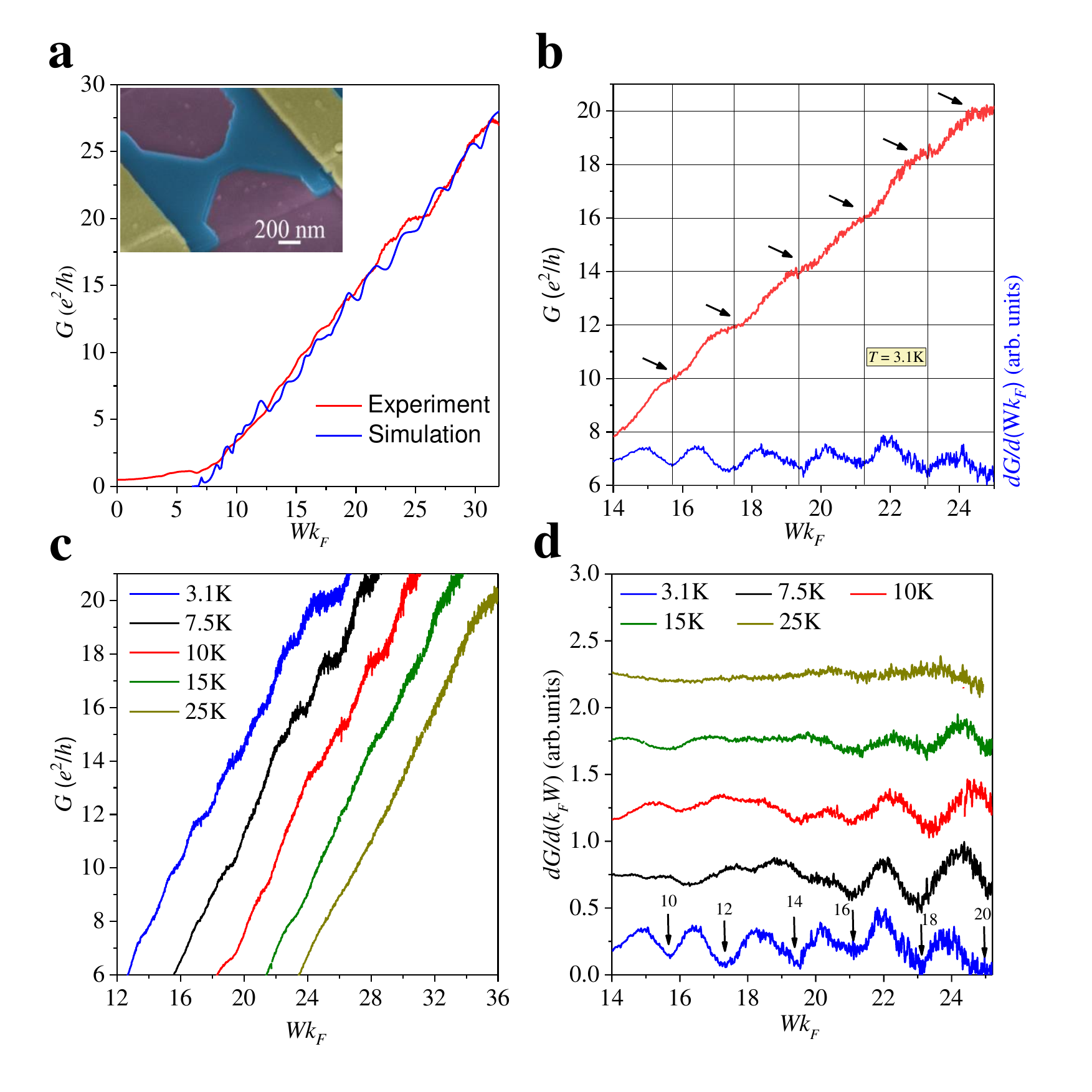}}
\caption{(a) Comparison of simulated and experimental conductance of a GNC of width $W=\SI{206}{\nano\meter}$. (b)~Conductance (red line) and transconductance (blue line) as a function $Wk_{F}$ measured at $T=3.1\,$K. Solid black arrows show the position of the plateaus of conductance separated by $2e^2/h$ and matching integer values from $G=10e^2/h$ onwards. (c)~Evolution of $G$ as a function of $Wk_{F}$ at different temperatures. The curves have been horizontally shifted by a factor $2.5Wk_{F}$ for clarity. (d)~Temperature dependence of the transconductance as a function of $Wk_{F}$. Black solid arrows represent the value of the plateaus of conductance in units of $e^2/h$ at $T=\SI{3.1}{\kelvin}$, matching the position of the minima in $dG/d(Wk_{F})$.}
\label{fig3}
\end{figure}

Figure~\ref{fig3}a shows the conductance $G$ as a function of $Wk_{F}$ measured at $T=\SI{3.1}{\kelvin}$ (red line) and the theoretical conductance (blue line) from our tight-binding simulations with the same edge roughness obtained from the AFM measurements (Figure~\ref{fig2}).

We note the good agreement of the experimental and simulated conductance as well as the linear dependence on $Wk_{F}$, characteristic of ballistic transport above a threshold value $Wk^{0}_{F}$. From the normalized gate voltage $\Delta V_{g}$  we can obtain the carrier density $n$ and  therefore the Fermi wavenumber $k_{F}=\sqrt{\pi n}$. Due to the presence of residual doping in our NC, we cannot tune the carrier density below a minimum value $n_0 \sim \SI{3.3e10}{\per\square\centi\meter}$ experimentally, and accordingly $Wk_F$ could not be tune below a threshold value  $Wk^{0}_{F} = \sqrt{\pi n_0} \, W \sim 6.7$. The transmission parameter $t$ is estimated from the linear fit of the  conductance when $Wk_{F}>Wk^{0}_{F}$ using Equation~(\ref{fit}). It is worth stressing the remarkable agreement of the slopes obtained from the theoretical and experimental conductance curves, yielding the transmission parameter $t=0.9$. This value is close to the fully transmitted carrier regime $t=1$, in contrast to the $t\sim 0.6$ given in Ref.~[\hspace*{-4px}\citenum{Terres}] for a graphene NC with a comparable width of $W=\SI{230}{\nano\meter}$ that was not defined by cryo-etching. For the electron-side ($V_g>V_g^{*}$), we also found a very similar transmission parameter $t=0.9$, as presented in the supplemental information S4. This result, supported by the remarkable agreement with theoretical calculations using realistic profiles of the edge roughness, is a clear indication of ballistic transport due to the very high mobility and smooth edges that arise from the novel use of the cryo-etching procedure in their preparation. The estimated value of $t$ in our structure is also significantly higher than the one obtained in much narrower NCs of width $W=\SI{100}{\nano\meter}$ produced by means of a HMDS treatment\cite{Caridad,Vito}. In these samples, low-roughness edges are generated mainly due to the fact that the etching process on a single layer graphene is more controllable and a clean edge on a single layer of material is easier to attain than the one obtained on a thicker encapsulated graphene nanostructure. The study of graphene NCs obtained by means of a HMDS treatment is also presented in the supplementary information for comparison (S5 and S6). In that case, the corresponding transmission parameters are significantly lower than the ones found in encapsulated graphene NCs defined by cryo-etching. 

A closer inspection of the conductance $G$ (red line) and the dimensionless transconductance $dG/d(Wk_{F})$ (blue line) versus $Wk_{F}$ is shown in Figure~\ref{fig3}b, signaling the second goal of this work. The plateaus of quantized conductance are clearly observed at regular intervals of $Wk_{F}$, corresponding to equally spaced marked minima of the transconductance. The quantization of $G$ is also very clear after reverting the direction of the current (see supplemental material S4). Up to six  plateaus appear at values of the conductance of $G=10,\, 12,\, 14,\, 16,\, 18,\, 20\, e^2/h$, hence the jump between neighbouring plateaus has been found to be $2e^{2}/h$. This value matches the theoretical estimation by Guimar\~{a}es \emph{et al.}, where the conductance of a graphene NC where $W\sim L$ is quantized into well-defined plateaus separated by $2e^{2}/h$\cite{Guimaraes}. Ihnatsenka \emph{et al.} put forward electron-electron interaction and boundary scattering as the major responsible for the appearance of plateaus separated by $2e^{2}/h$ instead of $4e^{2}/h$,\cite{Ihnatsenka} as it would be expected in ideal nanoribbons with perfectly smooth edges where both spin and valley degeneracy are preserved\cite{Kim}. To the best of our knowledge, this is the largest number of plateaus of conductance observed in single layer graphene samples that have been have been physically narrowed down into nanoconstriction-shape to date. While in bilayer graphene it is noticeably easier to confine carriers due to the presence of an already existing gap and hence plateaus of conductance are easier to achieve\cite{Overweg, HLee}, shaping electrostratically the conductive region into a constriction\cite{Overweg,Kim} allows to narrow-down the constriction further into a few tens of $\SI{}{\nano\meter}$ wide structure and to diminish the number of channels responsible of electric conduction. Our approach becomes a good compromise to achieve almost fully transmitted structures with smooth sidewalls where a systematic experimental study of NCs having differing widths may developed. In Figure~\ref{fig3}c we present the temperature dependence of the conductance (taken at $T=3.1,\,7.5,\,10,\,15$ and $25\SI{}{\kelvin}$), where the curves have been shifted by a factor of $2.5Wk_{F}$ for clarity. The presence of the plateaus is clear and they are noticeably visible below $T\sim 10\,$K. They start to smear out at higher temperatures until they are completely washed out at $T=25\,$K. Such a temperature dependence of $G$ can be better studied by inspecting the dimensionless transconductance as presented in Figure~\ref{fig3}d.  The amplitude of the periodic oscillations of the transconductance as a function of $Wk_{F}$ decreases by increasing temperature and finally disappear above $T=\SI{15}{\kelvin}$. A similar threshold temperature for the vanishing of the conductance quantization was very recently found in bilayer graphene constrictions\cite{HLee}. This behavior differs markedly from other non-desired effects, such as localized charge carriers, that disappear at higher temperature (see supplemental information S8).

In summary, we have proposed a novel implementation of the cryo-etching method, which we used for fabrication of high quality  graphene nanoconstrictions. The main advantage of the method is an unprecedented control of the edge definition and reduced roughness. The latter enabled us to fabricate high-mobility hBN-encapsulated graphene nanoconstrictions with well defined, sharp and extraordinarily smooth edges. These systems have been characterised by the standard AFM imaging techniques and then the obtained data on the realistic geometry was used in the large scale theoretical transport calculations. Results of such modelling agree very well with our experimental measurements and suggest that the transport in the system is almost fully ballistic. In particular, both the predicted and measured values of the transmission parameter are as high as $t=0.9$. On the other hand, the size quantization is clearly manifesting itself as a ladder of several equally $2e^2/h$-spaced plateaus of conductance at temperatures up to $T=10$ K, confirming the high quality of the samples.  We claim that the cryo-etching technique paves to study truly-ballistic-high-mobility graphene-based nanostructures where definition of the edges and the control of the scattering processes are crucial.


\begin{acknowledgement}

This research has been supported by  the Agencia Estatal de Investigaci\'on of Spain (Grants MAT2015-65274-R, MAT2016-75955 and TEC2015-65477-R), and the  Junta de Castilla y Le\'on (Grant SA256P18), including FEDER funds from the European Commission. V.~C. acknowledges N. Tombros and P. J. Zomer for introducing him in the transfer technique of two-dimensional materials. We are also thankful to D. L\'{o}pez and M. Vel\'{a}zquez for their kind help with Raman measurements. 

\end{acknowledgement}






\bibliography{references}

\providecommand{\latin}[1]{#1}
\makeatletter
\providecommand{\doi}
  {\begingroup\let\do\@makeother\dospecials
  \catcode`\{=1 \catcode`\}=2 \doi@aux}
\providecommand{\doi@aux}[1]{\endgroup\texttt{#1}}
\makeatother
\providecommand*\mcitethebibliography{\thebibliography}
\csname @ifundefined\endcsname{endmcitethebibliography}
  {\let\endmcitethebibliography\endthebibliography}{}
\begin{mcitethebibliography}{36}
\providecommand*\natexlab[1]{#1}
\providecommand*\mciteSetBstSublistMode[1]{}
\providecommand*\mciteSetBstMaxWidthForm[2]{}
\providecommand*\mciteBstWouldAddEndPuncttrue
  {\def\EndOfBibitem{\unskip.}}
\providecommand*\mciteBstWouldAddEndPunctfalse
  {\let\EndOfBibitem\relax}
\providecommand*\mciteSetBstMidEndSepPunct[3]{}
\providecommand*\mciteSetBstSublistLabelBeginEnd[3]{}
\providecommand*\EndOfBibitem{}
\mciteSetBstSublistMode{f}
\mciteSetBstMaxWidthForm{subitem}{(\alph{mcitesubitemcount})}
\mciteSetBstSublistLabelBeginEnd
  {\mcitemaxwidthsubitemform\space}
  {\relax}
  {\relax}

\bibitem[Han \latin{et~al.}(2007)Han, \"{O}zyilmaz, Zhang, and Kim]{Han}
Han,~M.~Y.; \"{O}zyilmaz,~B.; Zhang,~Y.; Kim,~P. \emph{Phys. Rev. Lett}
  \textbf{2007}, \emph{98 29}, 206805\relax
\mciteBstWouldAddEndPuncttrue
\mciteSetBstMidEndSepPunct{\mcitedefaultmidpunct}
{\mcitedefaultendpunct}{\mcitedefaultseppunct}\relax
\EndOfBibitem
\bibitem[van Wees \latin{et~al.}(1988)van Wees, van Houten, Beenakker,
  Williamson, Kouwenhoven, van~der Marel, and Foxon]{VanWees}
van Wees,~B.~J.; van Houten,~H.; Beenakker,~C.~W.~J.; Williamson,~J.~G.;
  Kouwenhoven,~L.~P.; van~der Marel,~D.; Foxon,~C.~T. \emph{Phys. Rev. Lett}
  \textbf{1988}, \emph{60}, 848\relax
\mciteBstWouldAddEndPuncttrue
\mciteSetBstMidEndSepPunct{\mcitedefaultmidpunct}
{\mcitedefaultendpunct}{\mcitedefaultseppunct}\relax
\EndOfBibitem
\bibitem[Caridad \latin{et~al.}(2018)Caridad, Calogero, Pedrinazzi, Santos,
  Impellizzeri, Gunst, Booth, Sordan, B{\o}ggild, and Brandbyge]{CaridadNL}
Caridad,~J.~M.; Calogero,~G.; Pedrinazzi,~P.; Santos,~J.~E.; Impellizzeri,~A.;
  Gunst,~T.; Booth,~T.~J.; Sordan,; B{\o}ggild,~P.; Brandbyge,~M. \emph{Nano
  Lett.} \textbf{2018}, \emph{8}, 4675\relax
\mciteBstWouldAddEndPuncttrue
\mciteSetBstMidEndSepPunct{\mcitedefaultmidpunct}
{\mcitedefaultendpunct}{\mcitedefaultseppunct}\relax
\EndOfBibitem
\bibitem[Kretinin \latin{et~al.}(2014)Kretinin, Cao, Tu, Yu, Jalil, Novoselov,
  Haigh, Gholinia, Mishchenko, Lozada, Georgiou, Woods, Withers, Blake, Eda,
  Wirsig, Hucho, Watanabe, Taniguchi, Geim, and V.]{Kretinin}
Kretinin,~A.~V. \latin{et~al.}  \emph{Nano Lett.} \textbf{2014}, \emph{14},
  3270\relax
\mciteBstWouldAddEndPuncttrue
\mciteSetBstMidEndSepPunct{\mcitedefaultmidpunct}
{\mcitedefaultendpunct}{\mcitedefaultseppunct}\relax
\EndOfBibitem
\bibitem[Kim \latin{et~al.}(2016)Kim, Choi, Lee, Watanabe, Taniguchi, Jhi, and
  Lee]{Kim}
Kim,~M.; Choi,~J.-H.; Lee,~S.-H.; Watanabe,~K.; Taniguchi,~T.; Jhi,~S.-H.;
  Lee,~H.-J. \emph{Nat. Phys.} \textbf{2016}, \emph{12}, 1022\relax
\mciteBstWouldAddEndPuncttrue
\mciteSetBstMidEndSepPunct{\mcitedefaultmidpunct}
{\mcitedefaultendpunct}{\mcitedefaultseppunct}\relax
\EndOfBibitem
\bibitem[Overweg \latin{et~al.}(2018)Overweg, Eggimann, Chen, Slizovskiy, Eich,
  Pisoni, Lee, Rickhaus, Watanabe, Taniguchi, Fal{\textquotesingle}ko, Ihn, and
  Ensslin]{Overweg}
Overweg,~H.; Eggimann,~H.; Chen,~X.; Slizovskiy,~S.; Eich,~M.; Pisoni,~R.;
  Lee,~Y.; Rickhaus,~P.; Watanabe,~K.; Taniguchi,~T.;
  Fal{\textquotesingle}ko,~V.; Ihn,~T.; Ensslin,~K. \emph{Nano Lett.}
  \textbf{2018}, \emph{18}, 553\relax
\mciteBstWouldAddEndPuncttrue
\mciteSetBstMidEndSepPunct{\mcitedefaultmidpunct}
{\mcitedefaultendpunct}{\mcitedefaultseppunct}\relax
\EndOfBibitem
\bibitem[Lu \latin{et~al.}(2010)Lu, Goldsmith, Strachan, Lim, Luo, and
  Charlie~Johnson]{Lu}
Lu,~Y.; Goldsmith,~B.; Strachan,~D.; Lim,~J.~H.; Luo,~Z.;
  Charlie~Johnson,~A.~T. \emph{Small} \textbf{2010}, \emph{6}, 2748\relax
\mciteBstWouldAddEndPuncttrue
\mciteSetBstMidEndSepPunct{\mcitedefaultmidpunct}
{\mcitedefaultendpunct}{\mcitedefaultseppunct}\relax
\EndOfBibitem
\bibitem[Lin \latin{et~al.}(2008)Lin, Perebeinos, Chen, and Avouris]{Lin}
Lin,~Y.-M.; Perebeinos,~V.; Chen,~Z.; Avouris,~P. \emph{Phys. Rev. B}
  \textbf{2008}, \emph{78}, 161409(R)\relax
\mciteBstWouldAddEndPuncttrue
\mciteSetBstMidEndSepPunct{\mcitedefaultmidpunct}
{\mcitedefaultendpunct}{\mcitedefaultseppunct}\relax
\EndOfBibitem
\bibitem[Lian \latin{et~al.}(2010)Lian, Tahy, Fang, Li, Xing, and Jena]{Lian}
Lian,~C.; Tahy,~K.; Fang,~T.; Li,~G.; Xing,~H.~G.; Jena,~D. \emph{Appl. Phys.
  Lett.} \textbf{2010}, \emph{96}, 103109\relax
\mciteBstWouldAddEndPuncttrue
\mciteSetBstMidEndSepPunct{\mcitedefaultmidpunct}
{\mcitedefaultendpunct}{\mcitedefaultseppunct}\relax
\EndOfBibitem
\bibitem[Tombros \latin{et~al.}(2011)Tombros, Veligura, Junesch, Guimar\~{a}es,
  Vera-Marun, T., and van Wees]{Niko}
Tombros,~N.; Veligura,~A.; Junesch,~J.; Guimar\~{a}es,~M.~H.~D.;
  Vera-Marun,~I.~J.; T.,~J.~H.; van Wees,~B. \emph{Nat. Phys.} \textbf{2011},
  \emph{7}, 697\relax
\mciteBstWouldAddEndPuncttrue
\mciteSetBstMidEndSepPunct{\mcitedefaultmidpunct}
{\mcitedefaultendpunct}{\mcitedefaultseppunct}\relax
\EndOfBibitem
\bibitem[Bischoff \latin{et~al.}(2015)Bischoff, Varlet, Simonet, Eich, Overweg,
  Ihn, and Ensslin]{Bischoff}
Bischoff,~D.; Varlet,~A.; Simonet,~P.; Eich,~M.; Overweg,~H.~C.; Ihn,~T.;
  Ensslin,~K. \emph{Appl. Phys. Rev.} \textbf{2015}, \emph{2}, 031301\relax
\mciteBstWouldAddEndPuncttrue
\mciteSetBstMidEndSepPunct{\mcitedefaultmidpunct}
{\mcitedefaultendpunct}{\mcitedefaultseppunct}\relax
\EndOfBibitem
\bibitem[Terr\'es \latin{et~al.}(2016)Terr\'es, Chizova, Libisch, Peiro,
  J\"orger, Engels, Girschik, Watanabe, Taniguchi, Rotkin, Burgd\"orfer, and
  Stampfer]{Terres}
Terr\'es,~B.; Chizova,~L.~A.; Libisch,~F.; Peiro,~J.; J\"orger,~D.; Engels,~S.;
  Girschik,~A.; Watanabe,~K.; Taniguchi,~T.; Rotkin,~S.~V.; Burgd\"orfer,~J.;
  Stampfer,~C. \emph{Nat. Commun.} \textbf{2016}, \emph{7}, 11528\relax
\mciteBstWouldAddEndPuncttrue
\mciteSetBstMidEndSepPunct{\mcitedefaultmidpunct}
{\mcitedefaultendpunct}{\mcitedefaultseppunct}\relax
\EndOfBibitem
\bibitem[Somanchi \latin{et~al.}(2017)Somanchi, Terr\'es, Peiro, Staggenborg,
  Watanabe, Taniguchi, Beschoten, and Stampfer]{Terres2}
Somanchi,~S.; Terr\'es,~B.; Peiro,~J.; Staggenborg,~M.; Watanabe,~K.;
  Taniguchi,~T.; Beschoten,~B.; Stampfer,~C. \emph{Ann. Phys. (Berlin)}
  \textbf{2017}, \emph{529}, 1700082\relax
\mciteBstWouldAddEndPuncttrue
\mciteSetBstMidEndSepPunct{\mcitedefaultmidpunct}
{\mcitedefaultendpunct}{\mcitedefaultseppunct}\relax
\EndOfBibitem
\bibitem[Caridad \latin{et~al.}(2018)Caridad, Power, Lotz, Shylau, Thomsen,
  Gammelgaard, Booth, Jauho, and B{\o}ggild]{Caridad}
Caridad,~J.~M.; Power,~S.~R.; Lotz,~M.~R.; Shylau,~A.~A.; Thomsen,~J.~D.;
  Gammelgaard,~L.; Booth,~T.~J.; Jauho,~A.-P.; B{\o}ggild,~P. \emph{Nat.
  Commun.} \textbf{2018}, \emph{9}, 659\relax
\mciteBstWouldAddEndPuncttrue
\mciteSetBstMidEndSepPunct{\mcitedefaultmidpunct}
{\mcitedefaultendpunct}{\mcitedefaultseppunct}\relax
\EndOfBibitem
\bibitem[Cleric\`{o} \latin{et~al.}(2018)Cleric\`{o}, Delgado~Notario,
  Saiz-Bret\'{i}n, {Hern\'{a}ndez Fuentevilla}, Malyshev, Lejarreta, Diez, and
  Dom\'{i}nguez-Adame]{Vito}
Cleric\`{o},~V.; Delgado~Notario,~J.~A.; Saiz-Bret\'{i}n,~M.; {Hern\'{a}ndez
  Fuentevilla},~C.; Malyshev,~A.~V.; Lejarreta,~J.~D.; Diez,~E.;
  Dom\'{i}nguez-Adame,~F. \emph{Phys. Status Solidi A} \textbf{2018},
  \emph{215}, 1701065\relax
\mciteBstWouldAddEndPuncttrue
\mciteSetBstMidEndSepPunct{\mcitedefaultmidpunct}
{\mcitedefaultendpunct}{\mcitedefaultseppunct}\relax
\EndOfBibitem
\bibitem[Defforgea \latin{et~al.}(2011)Defforgea, Songa, Gautiera, Tillocherc,
  Dussartc, S\'{e}bastien~Kouassi, and Tran-Van]{Defforgea}
Defforgea,~T.; Songa,~X.; Gautiera,~G.; Tillocherc,~T.; Dussartc,~R.;
  S\'{e}bastien~Kouassi,~S.; Tran-Van,~F. \emph{Sens. Actuators A Phys.}
  \textbf{2011}, \emph{170}, 114\relax
\mciteBstWouldAddEndPuncttrue
\mciteSetBstMidEndSepPunct{\mcitedefaultmidpunct}
{\mcitedefaultendpunct}{\mcitedefaultseppunct}\relax
\EndOfBibitem
\bibitem[Cadarso \latin{et~al.}(2017)Cadarso, Chidambaram, Jacot-Descombes, and
  Schift]{Cadarso}
Cadarso,~V.~J.; Chidambaram,~N.; Jacot-Descombes,~L.; Schift,~H.
  \emph{Microsyst. Nanoeng.} \textbf{2017}, \emph{3}, 17017\relax
\mciteBstWouldAddEndPuncttrue
\mciteSetBstMidEndSepPunct{\mcitedefaultmidpunct}
{\mcitedefaultendpunct}{\mcitedefaultseppunct}\relax
\EndOfBibitem
\bibitem[Liu \latin{et~al.}(2013)Liu, Wu, Harteneck, and Olynick]{LiuZ}
Liu,~Z.; Wu,~Y.; Harteneck,~B.; Olynick,~D. \emph{Nanotechnology}
  \textbf{2013}, \emph{24}, 015305\relax
\mciteBstWouldAddEndPuncttrue
\mciteSetBstMidEndSepPunct{\mcitedefaultmidpunct}
{\mcitedefaultendpunct}{\mcitedefaultseppunct}\relax
\EndOfBibitem
\bibitem[Pizzocchero \latin{et~al.}(2016)Pizzocchero, Gammelgaard, Jessen,
  Caridad, Wang, Hone, B{\o}ggild, and Booth]{Pizzocchero}
Pizzocchero,~F.; Gammelgaard,~L.; Jessen,~B.~S.; Caridad,~J.~M.; Wang,~L.;
  Hone,~J.; B{\o}ggild,~P.; Booth,~T.~J. \emph{Nat. Comm.} \textbf{2016},
  \emph{7}, 11894\relax
\mciteBstWouldAddEndPuncttrue
\mciteSetBstMidEndSepPunct{\mcitedefaultmidpunct}
{\mcitedefaultendpunct}{\mcitedefaultseppunct}\relax
\EndOfBibitem
\bibitem[Kim \latin{et~al.}(2009)Kim, Nah, Jo, Shahrjerdi, Colombo, Yao, Tutuc,
  and Banerjee]{Kim2}
Kim,~S.; Nah,~J.; Jo,~I.; Shahrjerdi,~D.; Colombo,~L.; Yao,~Z.; Tutuc,~E.;
  Banerjee,~S.~K. \emph{Appl. Phys. Lett.} \textbf{2009}, \emph{94},
  062107\relax
\mciteBstWouldAddEndPuncttrue
\mciteSetBstMidEndSepPunct{\mcitedefaultmidpunct}
{\mcitedefaultendpunct}{\mcitedefaultseppunct}\relax
\EndOfBibitem
\bibitem[Wallace(1947)]{Wallace47}
Wallace,~P.~R. \emph{Phys. Rev.} \textbf{1947}, \emph{71}, 622\relax
\mciteBstWouldAddEndPuncttrue
\mciteSetBstMidEndSepPunct{\mcitedefaultmidpunct}
{\mcitedefaultendpunct}{\mcitedefaultseppunct}\relax
\EndOfBibitem
\bibitem[{Castro Neto} \latin{et~al.}(2009){Castro Neto}, Guinea, Peres,
  Novoselov, and Geim]{CastroNeto09}
{Castro Neto},~A.~H.; Guinea,~F.; Peres,~N.~M.; Novoselov,~K.~S.; Geim,~A.~K.
  \emph{Rev. Mod. Phys.} \textbf{2009}, \emph{81}, 109\relax
\mciteBstWouldAddEndPuncttrue
\mciteSetBstMidEndSepPunct{\mcitedefaultmidpunct}
{\mcitedefaultendpunct}{\mcitedefaultseppunct}\relax
\EndOfBibitem
\bibitem[Reich \latin{et~al.}(2002)Reich, Maultzsch, and Thomsen]{Reich02}
Reich,~S.; Maultzsch,; Thomsen,~C. \emph{Physical Review B} \textbf{2002},
  \emph{66}, 035412\relax
\mciteBstWouldAddEndPuncttrue
\mciteSetBstMidEndSepPunct{\mcitedefaultmidpunct}
{\mcitedefaultendpunct}{\mcitedefaultseppunct}\relax
\EndOfBibitem
\bibitem[Morozov \latin{et~al.}(2008)Morozov, Novoselov, Katsnelson, Schedin,
  Elias, Jaszczak, and Geim]{Morozov08}
Morozov,~S.~V.; Novoselov,~K.~S.; Katsnelson,~M.~I.; Schedin,~F.; Elias,~D.~C.;
  Jaszczak,~J.~A.; Geim,~A.~K. \emph{Phys. Rev. Lett.} \textbf{2008},
  \emph{100}, 016602\relax
\mciteBstWouldAddEndPuncttrue
\mciteSetBstMidEndSepPunct{\mcitedefaultmidpunct}
{\mcitedefaultendpunct}{\mcitedefaultseppunct}\relax
\EndOfBibitem
\bibitem[Milo\u{s}evi\'{c} \latin{et~al.}(2010)Milo\u{s}evi\'{c}, Kep\u{c}ija,
  Dobard\u{z}i\'{c}, Damnjanovi\'{c}, Mohr, Maultzsch, and
  Thomsen]{Milosevic10}
Milo\u{s}evi\'{c},~I.; Kep\u{c}ija,~N.; Dobard\u{z}i\'{c},~E.;
  Damnjanovi\'{c},~M.; Mohr,~M.; Maultzsch,~J.; Thomsen,~C. \emph{Int. J. Mod.
  Phys. B} \textbf{2010}, \emph{24}, 655\relax
\mciteBstWouldAddEndPuncttrue
\mciteSetBstMidEndSepPunct{\mcitedefaultmidpunct}
{\mcitedefaultendpunct}{\mcitedefaultseppunct}\relax
\EndOfBibitem
\bibitem[Li \latin{et~al.}(2010)Li, Barry, Zavada, Nardelli, and Kim]{Li10}
Li,~X.; Barry,~E.~A.; Zavada,~J.~M.; Nardelli,~M.~B.; Kim,~K.~W. \emph{Appl.
  Phys. Lett.} \textbf{2010}, \emph{97}, 082101\relax
\mciteBstWouldAddEndPuncttrue
\mciteSetBstMidEndSepPunct{\mcitedefaultmidpunct}
{\mcitedefaultendpunct}{\mcitedefaultseppunct}\relax
\EndOfBibitem
\bibitem[Lent and Kirkner(1990)Lent, and Kirkner]{Lent90}
Lent,~C.~S.; Kirkner,~D.~J. \emph{J. Appl. Phys.} \textbf{1990}, \emph{67},
  6353\relax
\mciteBstWouldAddEndPuncttrue
\mciteSetBstMidEndSepPunct{\mcitedefaultmidpunct}
{\mcitedefaultendpunct}{\mcitedefaultseppunct}\relax
\EndOfBibitem
\bibitem[Mun\'{a}rriz(2014)]{Munarriz14}
Mun\'{a}rriz,~J. \emph{Modelling of plasmonic and graphene nanodevices};
  Springer: Berlin, 2014\relax
\mciteBstWouldAddEndPuncttrue
\mciteSetBstMidEndSepPunct{\mcitedefaultmidpunct}
{\mcitedefaultendpunct}{\mcitedefaultseppunct}\relax
\EndOfBibitem
\bibitem[Datta(1995)]{Datta95}
Datta,~S. \emph{Electronic transport in mesoscopic systems}; Cambridge
  University Press: Cambridge, 1995\relax
\mciteBstWouldAddEndPuncttrue
\mciteSetBstMidEndSepPunct{\mcitedefaultmidpunct}
{\mcitedefaultendpunct}{\mcitedefaultseppunct}\relax
\EndOfBibitem
\bibitem[Sharwin(1965)]{Sharwin}
Sharwin,~Y.~V. \emph{Sov. Phys. JETP} \textbf{1965}, \emph{21}, 655\relax
\mciteBstWouldAddEndPuncttrue
\mciteSetBstMidEndSepPunct{\mcitedefaultmidpunct}
{\mcitedefaultendpunct}{\mcitedefaultseppunct}\relax
\EndOfBibitem
\bibitem[de~Jong(1994)]{Jong}
de~Jong,~M.~J.~M. \emph{Phys. Rev. B} \textbf{1994}, \emph{49}, 7778\relax
\mciteBstWouldAddEndPuncttrue
\mciteSetBstMidEndSepPunct{\mcitedefaultmidpunct}
{\mcitedefaultendpunct}{\mcitedefaultseppunct}\relax
\EndOfBibitem
\bibitem[Kumar \latin{et~al.}(2017)Kumar, R.~Bandurin, Pellegrino, Cao,
  Principi, Guo, Auton, Shalom~Ben, Ponomarenko, Flakovich, Watanabe,
  Taniguchi, Grigorieva, Levitov, Polini, and Geim]{Kumar}
Kumar,~K. \latin{et~al.}  \emph{Nature Physics} \textbf{2017}, \emph{13},
  1182\relax
\mciteBstWouldAddEndPuncttrue
\mciteSetBstMidEndSepPunct{\mcitedefaultmidpunct}
{\mcitedefaultendpunct}{\mcitedefaultseppunct}\relax
\EndOfBibitem
\bibitem[Guimar\~{a}es \latin{et~al.}(2012)Guimar\~{a}es, Svevtsov, Waintal,
  and van Wees]{Guimaraes}
Guimar\~{a}es,~M.~H.~D.; Svevtsov,~O.; Waintal,~X.; van Wees,~B.~J. \emph{Phys.
  Rev. B} \textbf{2012}, \emph{85}, 075424\relax
\mciteBstWouldAddEndPuncttrue
\mciteSetBstMidEndSepPunct{\mcitedefaultmidpunct}
{\mcitedefaultendpunct}{\mcitedefaultseppunct}\relax
\EndOfBibitem
\bibitem[Ihnatsenka \latin{et~al.}(2012)Ihnatsenka, , and
  Kirczenow]{Ihnatsenka}
Ihnatsenka,~S.; ; Kirczenow,~G. \emph{Phys. Rev. B} \textbf{2012}, \emph{85},
  121407(R)\relax
\mciteBstWouldAddEndPuncttrue
\mciteSetBstMidEndSepPunct{\mcitedefaultmidpunct}
{\mcitedefaultendpunct}{\mcitedefaultseppunct}\relax
\EndOfBibitem
\bibitem[Lee \latin{et~al.}(2018)Lee, Park, Park, Lee, Watanabe, Taniguchi, and
  Lee]{HLee}
Lee,~H.; Park,~G.-H.; Park,~J.; Lee,~G.-H.; Watanabe,~K.; Taniguchi,~T.;
  Lee,~H. \emph{Nano Lett.} \textbf{2018}, \emph{18}, 5961\relax
\mciteBstWouldAddEndPuncttrue
\mciteSetBstMidEndSepPunct{\mcitedefaultmidpunct}
{\mcitedefaultendpunct}{\mcitedefaultseppunct}\relax
\EndOfBibitem
\end{mcitethebibliography}

\end{document}